\documentclass[10pt]{article}

\usepackage{epsfig, graphicx, lscape, amsmath, amsbsy, amstext, amsfonts}
\usepackage{array}

\usepackage{natbib}

\DeclareMathAlphabet{\mathb}{OT1}{cmr}{bx}{it}
\DeclareMathAlphabet{\mathsi}{OT1}{cmss}{m}{sl}
\DeclareMathAlphabet{\mathsb}{OT1}{cmss}{bx}{n}

\let\epsilon=\varepsilon
\let\theta=\vartheta
\let\phi=\varphi
\let\Phi=\varPhi
\let\Psi=\varPsi
\let\Omega=\varOmega

\newcommand{\reals}{\mathbb{R}}

\newcommand{\trans}{^{\scriptscriptstyle\mathsi{T}\unskip}}

\newtheorem{Thm}{Theorem}

\newcommand{\lcheck}[1]{\vphantom{|}_{{}_{#1}}\mspace{-1mu}|}
\newcommand{\rcheck}[1]{|_{{}_{#1}}}

\newcounter{thmlisti}
     {\begin{list}{\emph{(\roman{thmlisti})}}
          {\setlength{\topsep}{0pt}
           \setlength{\itemsep}{0pt}
           \setlength{\parsep}{0pt}
           \setlength{\labelsep}{0.5ex}
           \setlength{\labelwidth}{-2.8ex}
           \setlength{\leftmargin}{0ex}
           \usecounter{thmlisti}}}
     {\end{list}}

\begin{document}


\renewcommand{\baselinestretch}{1.2}



\renewcommand{\thefootnote}{}
$\ $\par


\fontsize{10.95}{14pt plus.8pt minus .6pt}\selectfont \vspace{0.8pc}
\centerline{\large\bf QUANTILE TOMOGRAPHY: USING QUANTILES WITH}
\vspace{2pt} \centerline{\large\bf
  MULTIVARIATE DATA}
\vspace{.4cm} \centerline{Linglong Kong and Ivan Mizera}
\vspace{.4cm} \centerline{\it Department of Mathematical and
Statistical Sciences, University of Alberta} \vspace{.55cm}
\fontsize{9}{11.5pt plus.8pt minus .6pt}\selectfont


\begin{quotation}
\noindent {\it Abstract:} The use of quantiles to obtain insights
about multivariate data is addressed. It is argued that incisive
insights can be obtained by considering directional quantiles, the
quantiles of projections. Directional quantile envelopes are proposed
as a way to condense this kind of information; it is demonstrated that
they are essentially halfspace (Tukey) depth levels sets, coinciding
for elliptic distributions (in particular multivariate normal) with
density contours. Relevant questions concerning their indexing, the
possibility of the reverse retrieval of directional quantile
information, invariance with respect to affine transformations, and
approximation/asymptotic properties are studied. It is argued that the
analysis in terms of directional quantiles and their envelopes offers
a straightforward probabilistic interpretation and thus conveys a
concrete quantitative meaning; the directional definition can be
adapted to elaborate frameworks, like estimation of extreme quantiles
and directional quantile regression, the regression of depth contours
on covariates. The latter facilitates the construction of multivariate
growth charts---the question that motivated all the development.

\vspace{9pt} \noindent {\it Key words and phrases:} Quantiles, Data
depth, Quantile regression, Growth charts. \par
\end{quotation}\par


\fontsize{10.95}{14pt plus.8pt minus .6pt}\selectfont

\par
\setcounter{equation}{0} 
\noindent {\bf 1. Introduction}

The concept of the quantile function is well rooted in the ordering of
$\reals$. For $0 < p <1$, the $p$-th \emph{quantile} (or
\emph{percentile}, if indexed by $100p$) of a probability distribution
$P$ is defined to be
\begin{equation*}
Q(p) = \inf \{ u \colon F(u) \geq p \},
\end{equation*}
where $F(u) = P((-\infty,u])$ is the cumulative distribution function
  of~$P$; see \citet{Eub86} or \citet{Sho00}. Essentially, $Q$ could
  be perceived as a function inverse to $F$; the more sophisticated
  definition is necessitated by a demand to treat formally cases when
  there is none, or more than one $q$ satisfying~$F(q) = p$. This is a
  well-known detail: if an alternative definition via the
  minimization of the integral
\begin{equation*}
\int \lcheck{p-1} x - q \rcheck{p} P(dx), \qquad
\text{where }
\lcheck{p-1} x \rcheck{p} =  x(p - I(x<0)),
\end{equation*}
is adopted, then the set, $\mathcal{Q}(p)$, of all minimizing $q$ may
be called, with \citet{Sho00}, a $p$-th \emph{quantile set} of~$P$;
and a prescription that returns, for all $p$, a unique element of this
(always nonempty, convex, and closed) set then constitutes a
\emph{quantile version}. \citet{HynFan96} review quantile versions
used in practice; these are implemented as options of the R function
\texttt{quantile} by \citet{FroHyn04}. While the ``inf'' version, as
defined above (not the default of \texttt{quantile}, but its option
``\verb|type=1|''), is preferred in theory and was used for all
pictures and computations in this paper, the practice often favors
other choices---like the ``midpoint'' version yielding the sample
median for $p=1/2$.

The potential of quantiles for blunt quantitative statements is well
known, and was noted already by classics: the reflection of Quetelet
was endorsed by \citet{Edg86, Edg93} and \citet{Gal88}. The
information, say, that $50$ is the $0.9$-th quantile leads to
unambiguous conclusion that about 10\% of the results are to be
expected beyond, and about 90\% below $50$. Compared to other
statistical uses---for which we refer to \cite{Par04} and the
references there---this ``descriptive grip'' is very palpable, and
hard to imagine beyond univariate context, in the multivariate
setting.

Yet, a natural and legitimate step in the analysis of multivariate
datasets is to apply quantiles to univariate functions of the original
data, the most immediate of such functions being projections. In
Section~2, we exemplify aspects of such exploration: in particular,
when projections in all directions are investigated simultaneously, we
observe a need for some kind of a summary, and propose in Section~3
``directional quantile envelopes'' to this end. The latter turn out to
be (if the ``inf'' quantile version is adhered to) level sets
(``contours'') of the halfspace (Tukey) depth---already a well-known
concept, whose directional interpretation is also hardly surprising;
the new name is thus feebly justified only by the fact that the exact
equality to depth contours does not hold true in general for other
quantile versions (a fact of rather formally mathematical than
data-analytic significance).

However, what we see as a potential contribution of this paper is
rather the observation that the directional interpretation of depth
contours not only gives them concrete probabilistic interpretation
(discussed in Sections~4 and 5) and thus quantitative meaning, but
that also enables to adapt them to more elaborate frameworks---like
estimation of extreme quantiles, and directional quantile regression.
In particular, directional interpretation makes possible, in a
simplest way, to regress depth contours on covariates (borrowing
strength as is typical in regression) and subsequently the
construction of bivariate growth charts---the methodology whose
pursuit was the original motivation for this paper. The applications
of the directional approach are introduced in Section~7, after the
discussion of some relevant properties in Section~6.

\medskip
\setcounter{equation}{0} 
\noindent {\bf 2. Quantile analysis of projections, illustrated on an
  example}

\begin{figure}[!b]
\begin{center}
\includegraphics[width=.45\textwidth, viewport=0 10 420 320, clip]{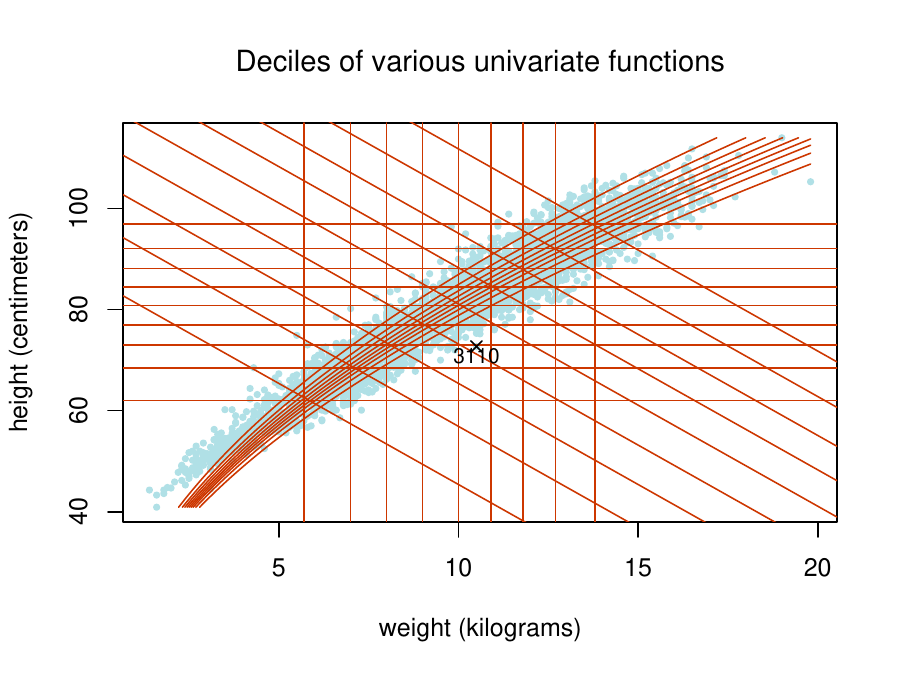}
\includegraphics[width=.45\textwidth, viewport=0 10 420 320, clip]{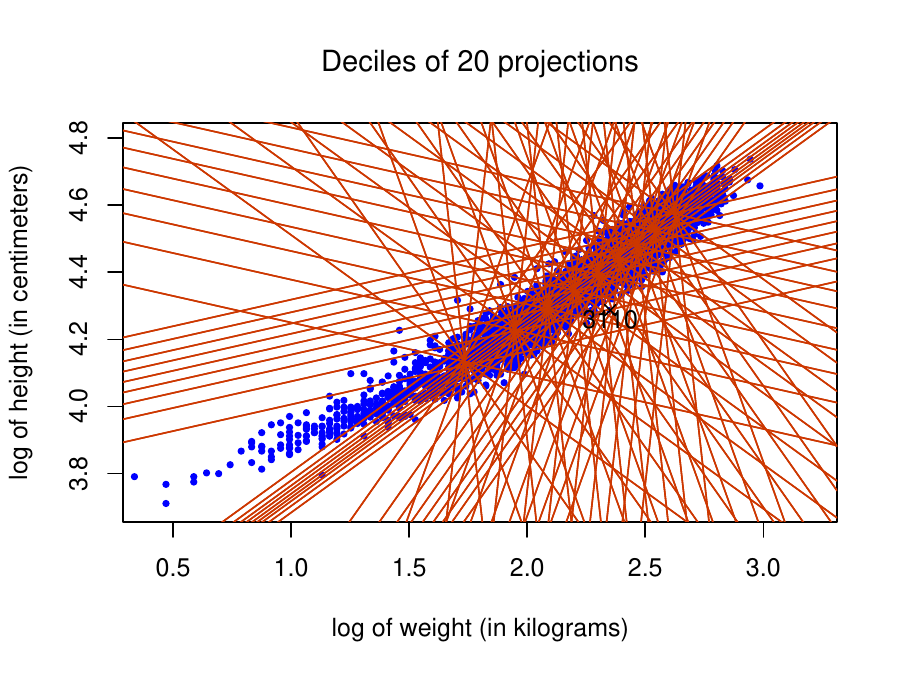}
\caption{\it Left panel: multivariate data typically offer insights
beyond the  marginal view, often through the quantiles of univariate
functions  of primary variables. Plotting the corresponding quantile
lines is  an appealing way to present this information. Right panel:
the plot gets quickly overloaded if multiple directions  and
indexing probabilities are requested. }
\label{f:1}
\end{center}
\end{figure}

Let us illustrate the possible objectives of the analysis using
quantiles in multivariate setting on an example. The left panel of
Figure~\ref{f:1} shows the scatterplot of the weight and height of
$4291$ Nepali children, aged between 3 and 60 months---the data
constituting a part of the Nepal Nutrition Intervention
Project-Sarlahi (NNIP-S, principal investigator Keith P. West, Jr.,
funded by the Agency of International Development). The horizontal and
vertical lines show the deciles of height and weight, respectively, of
the empirical distributions of the corresponding
variables---indicating the simple conclusions that can be reached
about the variables. For instance, the points above the upper
horizontal line correspond to $10$\% of the subjects exceeding the
others in height; similarly, the points right of the rightmost line
correspond to $10$\% of those exceeding the others in weight.

It would be interesting to know what proportion of the data
corresponds to the upper right corner, but this information is not
directly available (unless we count the points manually). Also,
regarding the subject labeled by $3110$, we can only say that its
weight is somewhat higher than, but otherwise fairly close to the
median; its height is about at the second decile, that is, exceeding
about $20$\% and exceeded by about $80$\% of its peers. Nevertheless,
one could argue that $3110$ is in certain sense extremal, outstanding
from the rest.

A possible way of substantiating this impression quantitatively is to
invoke Quetelet's \emph{body mass index} (hereafter \emph{BMI}),
defined as the ratio of weight to squared height (in the metric
system). The curved lines in the left panel of Figure~\ref{f:1} show
the deciles of the empirical distribution of the \emph{BMI}. We can
see that in terms of \emph{BMI}, the subject $3110$ is indeed extreme,
belonging to the group of $10$\% of those with maximal \emph{BMI}.

An expert on nutrition may dispute the relevance of \emph{BMI} for
young children, and remind us of possible alternatives---for instance,
the Rohrer index (ratio of weight to \emph{cubed} height, hereafter
\emph{ROI}). However, we do not think that the problem lies in
deciding whether that or another index is to be preferred; the essence
of the data may lie well beyond the index-style of description. For
example, suppose that we would like to make quantitative statements
about the subjects represented by the points in the upper right and
lower left rectangles. Since we are not aware of any relevant index
related to this objective, we may simply look, in the left panel of
Figure~\ref{f:1}, at the deciles of some suitable linear combination
of weight and height.

Pursuing vague objectives in nonlinear realm may be hard---there are
simply too many choices. A possible solution is to limit the attention
only to \emph{linear} functions of the original data; note that the
``\emph{BMI} contours'' in the left panel of Figure~\ref{f:1} are not
that badly approximated by straight lines. We can do even better by
taking the \emph{logarithms} of weight and height as primary
variables---then we can investigate both \emph{BMI} and \emph{ROI}
among their linear combinations, and possibly much more. Therefore, we
switch to the logarithmic scale, starting from the right panel of
Figure~\ref{f:1}.

Rather than this technical detail, however, the more important outcome
of our exploration is that quantiles of certain functions of variables
(in particular, linear combinations) may provide valuable information
about multivariate data. Focusing on linear combinations, we realize
that it is sufficient to look exclusively at projections; any other
linear combination is a multiple of a projection, and the quantile of
a multiple is the multiple of the quantile. In other words, we believe
that insights about data can be obtained by looking at the directional
quantiles.

The right panel of Figure~\ref{f:1} thus shows the plot of the
logarithms of weight and height, together with superimposed lines
indicating deciles in $20$ uniformly spaced directions. While these
\emph{directional quantile lines} are an appealing way to present the
directional quantile information, we have to admit that the plot
becomes quickly overloaded if multiple directions and indexing
probabilities are requested. (While our focus here is not exclusively
graphical, the task of plotting is probably the most palpable one to
epitomize our objectives.) Therefore, we would like to achieve some
compression of the directional quantile information; to this end, we
propose \emph{directional quantile envelopes}.



\medskip
\setcounter{equation}{0} 
\noindent {\bf 3. Directional quantile envelopes and halfspace depth}

Notationally, it is often convenient to work with random variables or
vectors, and write
\begin{equation*}
Q(p) = Q(p, X) = \inf \{ u \colon \mathbb{P}[X \leq u] \geq p \},
\end{equation*}
despite that the quantiles depend only on the distribution, $P$, of
$X$. Hereafter, $X$ will always stand for a random vector with the
distribution $P$; the apparent notational convention is to suppress
the dependence on $X$ when no confusion may arise. We call any vector
with unit norm in $\reals^d$ a \emph{normalized direction}, and denote
the set of all such vectors by $\mathbb{S}^{d-1}$. Given a normalized
direction $s \in \mathbb{S}^{d-1}$ and $0< p < 1$, the $p$-th
\emph{directional quantile}, in the direction $s$, is nothing but the
$p$-th quantile of the corresponding projection of the distribution
of~$X$,
\begin{equation*}
Q(p,s) = Q(p,s,X) = Q(p, s\trans X).
\end{equation*}
A related notion is the $p$-th \emph{directional quantile hyperplane},
given by the equation $s\trans x = Q(p,s)$. For $d=2$, the hyperplanes
amount to lines---which in our figures indicate how directional
quantiles divide the data.

The $p$-th directional quantile in the direction $s$ and the
$($$1$$-$$p$$)$-th directional quantile in the direction $-s$ are not
necessarily equal---due to the $\inf$ convention employed in their
definition. Nonetheless, they often coincide---for instance, it is not
possible to distinguish between any $p$-th and $($$1$$-$$p$$)$-th directional
quantile hyperplanes if for any projection of $P$, all quantile sets
are singletons. A sufficient condition for this is that $P$ has
\emph{contiguous support}: there is no intersection of halfspaces with
parallel boundaries that has nonempty interior but zero probability
$P$ and divides the support of $P$ to two parts. (Note that if the
support is not contiguous, it is not connected; however, it may be
disconnected and still contiguous.) We believe that contiguous support
is a fairly typical virtue of population distributions, and
consequently will limit most of our attention to $p$ from $(0, 1/2]$.
%

\begin{figure}[!b]
\begin{center}
\includegraphics[width=.445\textwidth, viewport=0 10 420 310, clip]{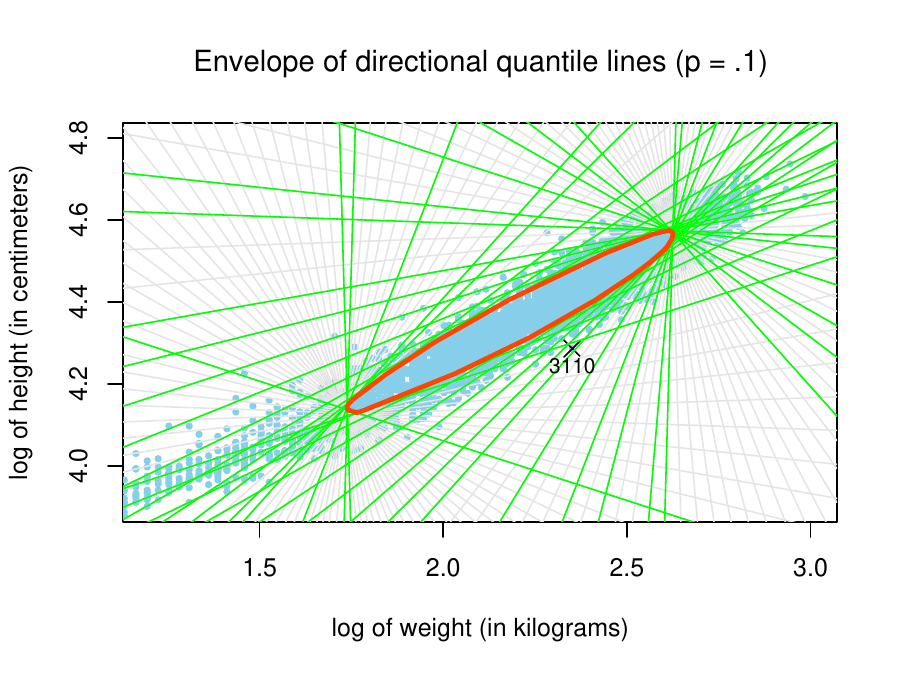}
\includegraphics[width=.445\textwidth, viewport=0 10 420 310, clip]{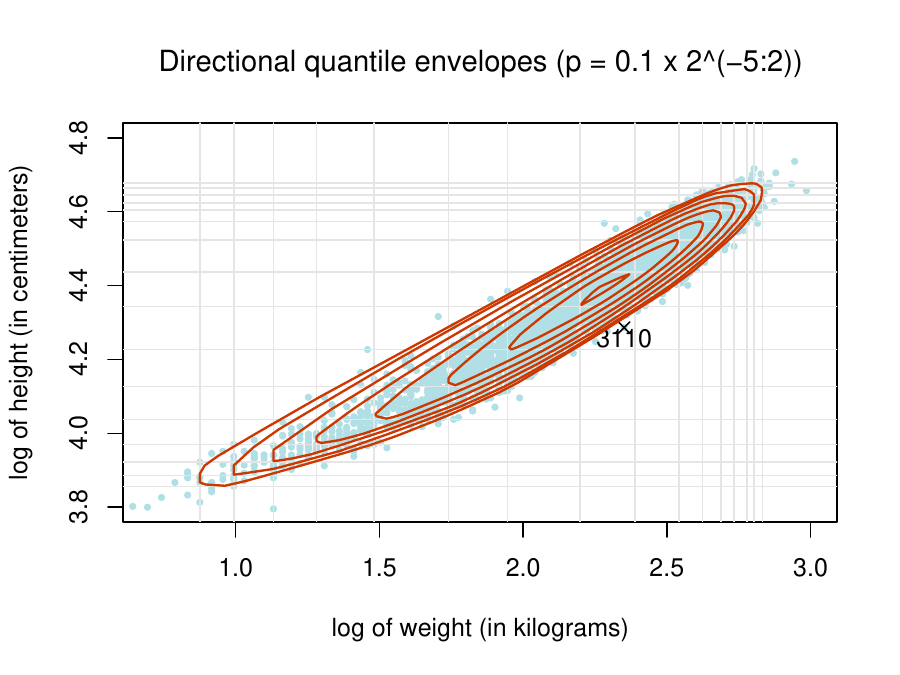}
\caption{\it Left panel: for fixed $p$, we form the inner envelope of
  the directional quantile lines. Right panel: directional quantile
  envelopes for $p = 2^i/10$, $i=-5,\dots,2$. In the central part, the
  contours resemble those obtained by fitting normal distribution; in
  the tail area, they adapt more to the specific shape of the data.
  Several $p$ can be accommodated simultaneously, and the directional
  quantile information can be retrieved from the contours in 
  a relatively straightforward retrieval way.}
\label{f:7}
\end{center}
\end{figure}

Contiguity of the support of $X$ is also one of the things that
implies continuity, in $s$, of the directional quantiles. The
following theorem is formulated slightly more generally, to allow for
alternative quantile versions and later asymptotic considerations.
Using the theorem with $X_n = X$ shows that the directional quantiles
depend continuously on $s$ for all empirical, and many population
distributions.
\begin{Thm}
\label{t:quacont} If the support of $X$ is bounded, then
$\mathcal{Q}(p,s)$ is a continuous function of $s$, for every $p \in
(0,1)$. The same holds true when the support of $X$ is contiguous;
moreover, if a sequence of random vectors $X_n$ converges almost
surely to $X$, and $s_n \to s$, then $\mathcal{Q}(p,s_n,X_n)$
converges to $\mathcal{Q}(p,s,X)$ in the Pompeiu-Hausdorff distance,
for every $p \in (0,1)$.
\end{Thm}
The terminology of ``Pompeiu-Hausdorff'' is that of \citet{RocWet98}.

The idea of what constitutes the inner envelope of the directional
quantile hyperplanes is quite clear from the left panel of
Figure~\ref{f:7}. More formally, for $p \in (0,1/2]$, the $p$-th
  \emph{directional quantile envelope} generated by $Q(p,s)$ is
  defined as the intersection,
\begin{equation*}
D(p) = \bigcap_{s \in \mathbb{S}^{d-1}} H(s,Q(p,s)),
\end{equation*}
where $H(s,q) = \{ x \colon s\trans x \geq q \}$ is the
\emph{supporting halfspace} determined by $s\in \mathbb{S}^{d-1}$ and
$q \in \reals$. In case the intersection will be taken only over a
subset $A \subseteq \mathbb{S}^{d-1}$ of all possible directions (for
instance, in the numerical construction of the envelopes), we will
write $D_A(p)$; in the spirit of this notation, $D(p) =
D_{\mathbb{S}^{d-1}}(p)$. Directional quantile envelopes are convex
(being intersections of convex sets) and bounded (for $D_A(p)$, this
is true whenever $A$ is not contained in any closed halfspace whose
boundary contains the origin). They have a close connection to what is
known as (halfspace or Tukey) depth, first considered by
\citet{Hod55}; \citet{Tuk75} proposed depth contours for plotting
bivariate data, in a spirit close to ours. Let $P$ be a distribution
in $\reals^d$. Recall that the \emph{depth}, $d(x)$, of a point $x \in
\reals^d$, is defined as $\inf P(H)$, where $H$ runs over all closed
halfspaces containing $x$ (or, equivalently, over all closed
halfspaces with $x$ lying on their boundary).
\begin{Thm}
\label{t:basic} For every $p \in (0, 1/2]$, the directional quantile
envelope is equal
  to the upper level set of depth: $D(p) = \{ x \colon d(x) \geq p
  \}$.
\end{Thm}
Theorem~\ref{t:basic} implies that directional quantile envelopes are
nonempty for $p \leq 1/(d+1)$, in the two-dimensional case for~$p \leq
1/3$, due to a result known as a centerpoint theorem---see
\citet{DonGas92} or \citet{Miz02}.

We remark that Theorem~\ref{t:basic} is rigorously true only for the
``inf'' version of the quantile definition. In practice, some other
version may be preferred, for instance, to allow for constructing
contours interpolating between various depth level sets. Most of the
other theorems in this paper hold true also for other versions, as can
be seen in the Appendix; this fact gives some justification for
calling what are essentially ``depth contours'' by a new name
``directional quantile envelopes''. All interpolated versions of
quantiles yield somewhat smaller envelopes; \citet{RouRut99} point out
that this is also the case for the related notion of halfspace trimmed
contours of \citet{MasThe94}. These subtle differences vanish in
regular situations---for instance, for absolutely continuous
distributions with positive densities.

\medskip
\setcounter{equation}{0} 
\noindent {\bf 4. Indexing, illustrated on the multivariate normal
  distribution}

As can be seen on the right panel of Figure~\ref{f:7}, suppressing the
underlying directional quantile lines (still shown in the left panel
of Figure~\ref{f:7}) allows for accommodating several~$p$
simultaneously. In the central part, the contours have elliptical
shape, resembling the density contours of the multivariate normal
distribution. Indeed, Theorem~\ref{t:elliptic} below implies that
directional quantile envelopes coincide with the density contours for
any elliptic distribution---in particular, for the multivariate
normal. In such a context, an intriguing question of practical
importance is that of indexing: which particular contours of the
fitted normal distribution would correspond to which~$p$?

\begin{figure}[!b]
\begin{center}
\includegraphics[width=.48\textwidth, viewport=0 10 420 310, clip]{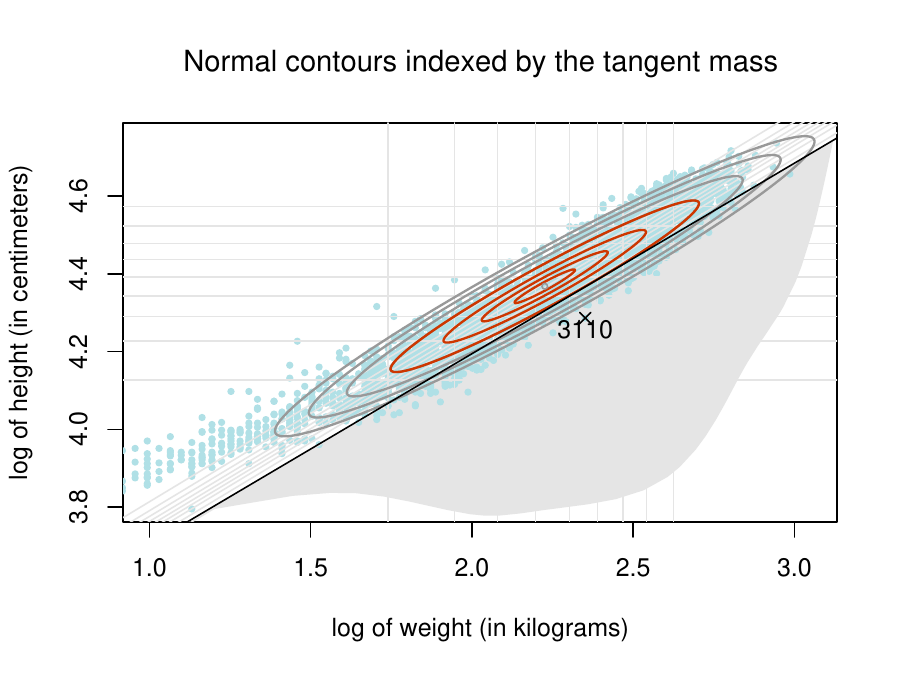}
\includegraphics[width=.48\textwidth, viewport=0 10 420 310, clip]{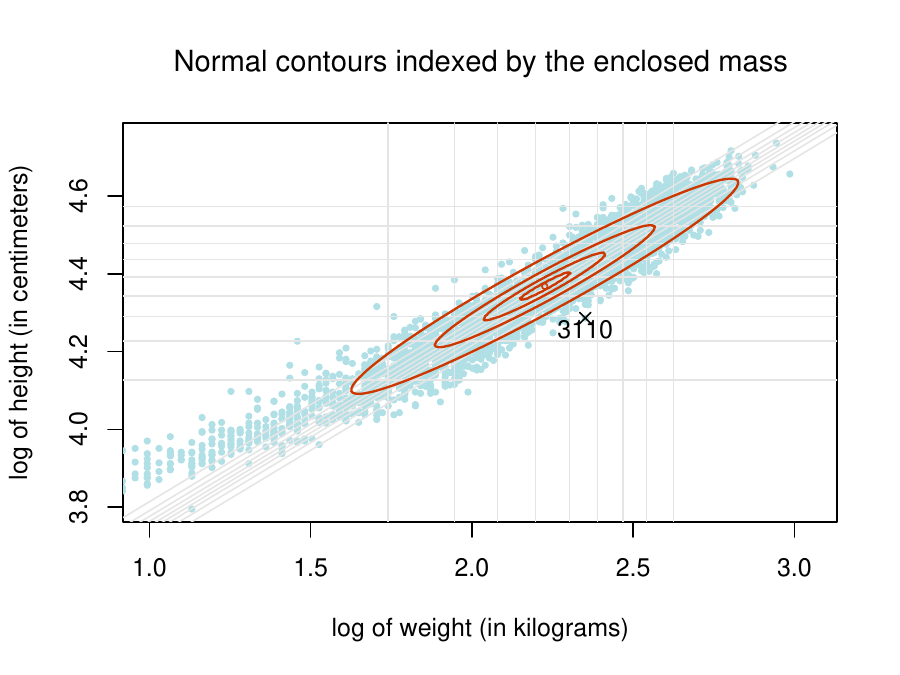}
\caption{\it Left panel: if indexed by the tangent mass, the contours
  of the fitted normal distribution theoretically match projected
  quantiles. The halfplane tangent to the contour and passing through
  the point contains exactly $p$ of the mass of the fitted
  multivariate normal distribution. Right panel: if indexed by the
  mass they enclose, the contours of the fitted normal distribution do
  not interact well with directional and marginal quantiles.}
  \label{f:5}
\end{center}
\end{figure}

Our definition of directional quantile envelopes leads to what we call
``indexing by the tangent mass''. It is illustrated by the grey
shading in the right panel of Figure~\ref{f:5}: given any point of the
contour, the halfplane passing through the point and tangent to the
contour contains exactly $p$ of the mass of the fitted multivariate
normal distribution. This extrapolates the univariate fact that $p$-th
and $($$1$$-$$p$$)$-th quantiles mark the boundaries of the halfspaces
containing exactly $p$ of the distribution mass. For the standard
normal distribution, the contour corresponding to $p$ is that matching
the univariate quantiles indexed by $p$ and $1$$-$$p$ when projected on
the coordinate axes, and can be found by transforming to the standard
form and the subsequent inverse transformation.

Thinking in terms of elliptic confidence sets may suggest another
alternative, ``indexing by the enclosed mass'', which generalizes the
univariate fact that the $p$-th and $($$1$$-$$p$$)$-th quantiles
together leave $2p$ of the distribution mass outside their convex
hull. In such a case, the contours corresponding to deciles would be
those enclosing $0.8$, $0.6$, $0.4$, $0.2$ of the mass of the fitted
normal distribution, together with the contour consisting of the
single point located at the mode.
This type of indexing can be seen in the right panel of
Figure~\ref{f:5}; this shows now that the subject represented by the
point 3110 lies in the outstanding 20\% of the sample; however, this
exceptionality is somewhat ``generic''---expressed not only through
the company of similar subjects with large weight given the height,
but also by the company of those with small weight given the height,
and of those with small height and weight altogether.

Contrary to that, the 10\% extremality of 3110 suggested by the left
panel can be interpreted as substantial: it is carried by the company
of subjects with similar nature, those with large weight given the
height. Note that the boundary of the greyed halfspace is almost
identical with the line indicating the $(0.9)$-th quantile of the
\emph{BMI}; hence the picture shows that in this case, the extremality
of 3110 may be interpreted in terms of \emph{BMI}. Also, unlike the
indexing by the enclosed mass, indexing by the tangent mass interacts
well with marginal and directional quantiles.



Methods based on fitting normal distribution are still somewhat
central to multivariate statistics; we find it thus encouraging that
for the normal distribution, our proposed contours coincide with the
contours of its density, the objects that \citet{Eva82} defines, in
one of the first papers on the subject, to be ``bivariate quantiles''.
However, the approach based on fitting normal distribution would have
all virtues of an ideal, if the hypothesized distribution would be
``closely followed'' by the data---as for our example occurs in the
central part, as can be seen in the right panel of Figure~\ref{f:7},
but not that much on the fringe of the data cloud. The elliptic
contours would be considerably off there; but we can see that the
directional quantile envelopes adapt to the specific shape of the
data, and thus behave rather in a nonparametric way.

\medskip
\setcounter{equation}{0} 
\noindent {\bf 5. Recovery of directional quantile information}

Obviously, directional quantile envelopes suppress some information
contained in directional quantiles; a question of paramount importance
is how far it is possible to get this information back. Let $e$ be a
point lying on the boundary, $\partial E$, of a bounded convex set $E
\subset \reals^d$. A \emph{tangent} of $E$ at $e$ is any hyperplane
(line) containing $e$ that has empty intersection with the interior of
$E$. Such a line determines the corresponding \emph{tangent
  halfspace}, the halfspace that has the tangent as its boundary and
its interior does not contain any point of $E$. The \emph{maximal mass
  at a hyperplane} is defined as $ \Delta(P) = \sup \{
\mathbb{P}[s\trans X=c] \colon s \in \mathbb{S}^{d-1},\; c\in
\mathbb{R}\}. $ The following theorem provides a practical guideline
for recovering the directional quantile information and is thus
essential in interpreting directional quantile envelopes.

\begin{figure}[!b]
\begin{center}
\includegraphics[width=.47\textwidth, height=.17\textheight,viewport=50 60 330 300, clip]{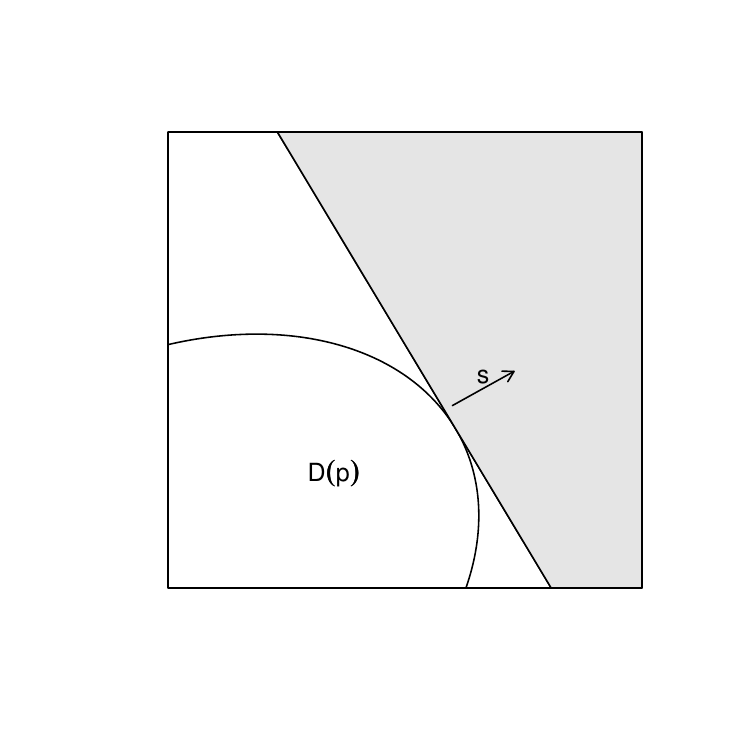}%
\hspace*{18pt}
\includegraphics[width=.47\textwidth, height=.17\textheight,viewport=50 60 330 300, clip]{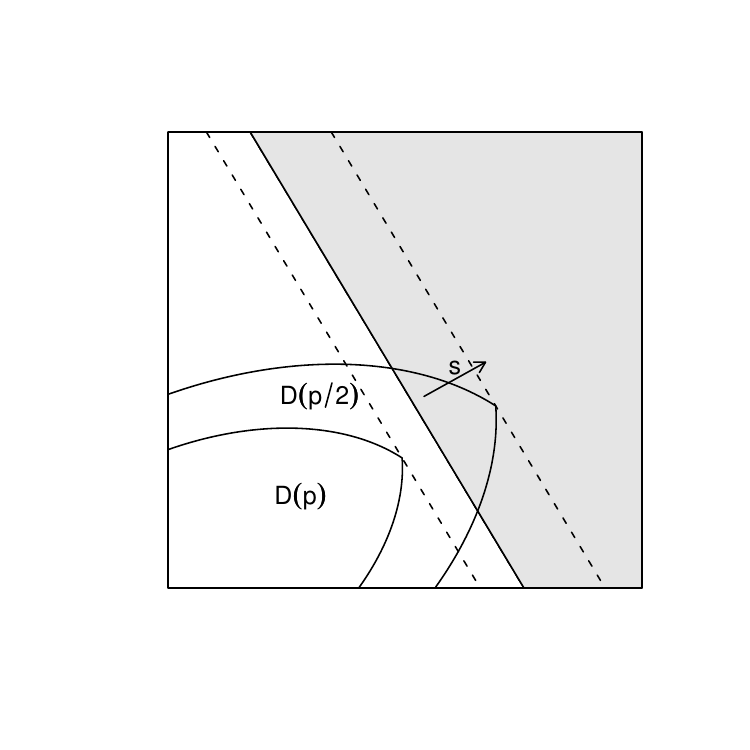}
\caption{\it Left panel: if the tangent line to the $p$-th
directional
  quantile envelope is unique, then the tangential halfspace is the
  $p$-th directional quantile halfspace, in the given direction. Right
  panel: if the tangent line is nonunique, then this directional
  quantile halfspace lies between $p$-th and $(p/2)$-th directional
  quantile envelope.}
\label{f:9}
\end{center}
\end{figure}

\begin{Thm}
\label{t:linglong} Let $P$ be a distribution in $\reals^d$, and let
$p \in (0,1/2]$. If
  $H$ is a tangent halfspace of $D(p)$, then $p\leq P(H)\leq
  2p+\Delta(P)$. Moreover, $p\leq P(H)\leq p+\Delta(P)$, if $\partial H$
  is the unique tangent of $D(p)$ at some point from $H\cap \partial
  D(p)$; in particular, $P(H)=p$ if $\Delta(P)=0$.

If $A\in \mathbb{S}^{d-1}$ is a finite set of directions and $H$ is
a tangent halfspace of $D_A(p)$, then still $P(H)\leq 2p+\Delta(P)$,
and $p \leq P(H)\leq p+\Delta(P)$, if $\partial H$ is the unique
tangent of
$D_A(p)$ at \emph{some} point from $H \cap \partial D(p)$. In
particular, $P(H)=p$ if $\Delta(P)=0$.
\end{Thm}
The left panel of Figure~\ref{f:9} shows the situation when the
tangent to the directional quantile envelope is unique. For a
population distribution with $\Delta(P) = 0$, one can uniquely
identify the directional quantile in the direction perpendicular to
the tangent. The visual determination of the uniqueness of the tangent
may be slightly in the eye of beholder; if this is undesirable, then
one may switch to a strictly finite-sample viewpoint, in which the
directional quantile envelopes of empirical probability distributions
are polygons and the uniquely identifiable directional quantile lines
are those that contain a boundary segment of the polygon.
%
%
If the tangent is not unique, the situation shown in the right panel
of Figure~\ref{f:9}, then the exact identification of the directional
quantile line is not possible; nevertheless, the inequality $P(H) \leq
2p$ given by Theorem~\ref{t:linglong} allows at least for its
approximate localization (especially when the plotted envelopes are so
chosen that $p$ follows a geometric progression with multiplier $1/2$,
as in the right panel of Figure~\ref{f:7}; note that such choice gives
approximately equispaced contours for normal distribution in the tail
area).

A boundary point of a convex set that admits more than one tangent is
called \emph{rough} (singular). It is known---see Theorem 2.2.4 of
\citet{Sch93}---that such points are quite exceptional; in particular,
for any closed convex set in $\reals^2$, the set of rough points is at
most countable. Convex, closed subsets of $\reals^d$ having no rough
points are called \emph{smooth}, consistently with the natural
geometric perception of the boundary in this case. If $D(p)$ is
smooth, then the collection of its tangent halfspaces is in one-one
correspondence with the collection of $p$-th directional quantile
halfspaces, with the same boundaries, but in opposite directions.

Although the assumption of smoothness may sound optimistically mild,
the examples in \citet{RouRut99} show that distributions with depth
contours having a few rough points are not that uncommon. However, it
may be argued that all these examples have somewhat contrived flavor,
especially when the support of the distribution is some regular
geometric figure; it is not unlikely that typical population
distributions have smooth depth contours---but we were not able to
find a suitable formal condition reinforcing this belief, beyond the
somewhat restricted realm of elliptically-contoured distributions.
Recall that the distribution is called \emph{elliptic} if it can be
transformed by an affine transformation to a circularly symmetric,
rotationally-invariant distribution. The following theorem, in
particular, confirms the fact mentioned earlier: normal contours allow
for the retrieval of \emph{all} directional quantile lines.
\begin{Thm}
\label{t:elliptic} The directional quantile envelopes of any
elliptic distribution are smooth.
\end{Thm}

Even if the tangent line at a boundary point of a directional quantile
envelope is nonunique, it does not necessarily mean that the
information about certain directional quantiles is lost. Although the
directional quantile is not retrievable from the envelope
\emph{directly}, in a straightforward manner, it may be possible to
reconstruct it from the totality of all envelopes. Formally this means
that the collection of directional quantile envelopes determines the
distribution uniquely.
Surprisingly, this plausible property has not yet been rigorously
proved in full generality, positive answers have been established only
for partial cases: depth functions uniquely characterize empirical
\citep{StrRou99}, and more generally atomic \citep{Kos02}
distributions, and also absolutely continuous distributions with
compact support \citep{Kos01}. A small progress in this direction is
the following result of \citet{KongZuo10} concerning distributions
with smooth depth contours.
%
\begin{Thm}
\label{t:smooth} If the directional quantile envelopes $D(p)$, of a
probability distribution $P$ in $\reals^d$ with contiguous support,
have smooth boundaries for every $p \in (0,1/2)$, then there is no
other probability distribution with the same directional quantile
envelopes.
\end{Thm}

\medskip
\setcounter{equation}{0} 
\noindent {\bf 6. Invariance, approximation, and estimation}

Our considerations so far bore rather a probabilistic than a
statistical character; to advance the latter, the most straightforward
way is to invoke the principle called ``na\"ive statistics'' by
\citet{HajVor77}, or ``analogy'' by \citet{Gol68} and \citet{Man88},
or ``plug-in principle'' by \citet{EfrTib93}: that is, to apply the
general definition to empirical distributions.

From the general point of view, we are interested in the
\emph{population quantile information}, directional quantiles of some
\emph{population distribution}; we believe that our data come, in some
sampling manner, from this distribution. To facilitate theoretical
analysis of typical cases, it is often reasonable to posit some
assumptions on this distribution; while a membership in a parametric
family, or ellipticity may be considered too stringent, continuity
assumptions are often acceptable. Our general strategy is to estimate
the result of the evaluation of a functional on the population
distribution via the application of the same functional to the
empirical distribution supported by the data: more specifically, we
want to estimate, for fixed $p$, the directional quantiles $Q(p,s)$ by
$\hat{Q}(p,s)$, and then use these estimates to generate the estimated
directional quantile envelope.

Recall that an operator (we use this word to indicate that unlike a
``function'', an ``operator'' can be set-valued) assigning a point or
a set $T$ in $\reals^d$ to a collection of datapoints $x_i \in
\reals^d$, is called \emph{affine equivariant}, if its value is $BT+b$
when evaluated from the datapoints $Bx_i+b$, for any nonsingular
matrix $B$ and any $b \in \reals^d$. (If $T$ is a set, then the
transformations are performed elementwise.) By Theorem~\ref{t:basic},
the estimated and population directional quantile envelopes are the
level sets of depth applied to the empirical and population
distributions, respectively; the properties of depth then imply their
affine equivariance. When directional quantiles are estimated by some
other means, for instance as a response of a quantile regression, the
affine equivariance of the resulting envelopes may be not that clear.
Nevertheless, the affine equivariance still takes place under mild
assumptions on the directional quantile estimators. Recall that an
operator that assigns a point, or set of points, $T$, in $\reals$, to
a random variable $X$ is called \emph{translation equivariant}, if its
value for $X+b$ coincides with $T+b$, and \emph{scale equivariant}, if
its value for $cX$ coincides with $cT$.
It is a direct consequence of the definition that the directional
quantile operator $Q(p,\cdot)$ is translation and scale equivariant,
for any $0<p<1$ (and for the ``inf'' version; for every other version,
the equivariance has to be checked individually---usually a
straightforward task).

\begin{Thm}
\label{t:affine} Suppose that directional quantile estimators
$\hat{Q}(p,s)$ are translation and scale equivariant, for all $s \in
\mathbb{S}^{d-1}$ and fixed $p$. Then the directional quantile
envelope generated by these estimators is affine equivariant.
\end{Thm}

The next question investigated in this section is how quantile
directional envelopes behave when not determined exactly, but only
approximately. There are several reasons to investigate that. The
numerical motivation stems from the fact that in practice we are not
able to take all directions to construct the directional quantile
envelope---what is really constructed is rather an approximate
envelope $D_A(p)$, and we are interested in the quality of this
approximation. Obviously, $D(p) \subseteq D_A(p)$; moreover, we
believe that a decent collection, $A$, of directions that reasonably
fill $\mathbb{S}^{d-1}$ makes the approximation quite satisfactory.
While the experimental evidence does not contradict this belief---for
the left panel of Figure~\ref{f:7} we used only $100$ uniformly spaced
directions, for the right panel of Figure~\ref{f:7} we took $1009$,
and hardly any difference can be seen for $p=0.1$---some theoretical
support would be desirable too. The statistical motivation for the
investigation of the approximation effects comes from the fact that
our directional quantiles are typically not the ``true'', but
``estimated'' ones. If we believe that this estimation is
consistent---we can show, in some customary probabilistic framework,
that estimates become more and more precise, say, with growing sample
size---then we are again interested whether this consistent behavior
of individual directional quantiles translates into something
analogous for their envelopes.
%
%
\begin{Thm}
\label{t:hausdorff} Suppose that $A_1 \subseteq A_2 \subseteq A_3
\subseteq \dots$ is a sequence of closed sets with its union dense
in a closed set $A \subseteq \mathbb{S}^{d-1}$, not contained in any
closed halfspace whose boundary contains the origin. If,
for every sequence $s_n \in A_n$ that converges to $s \in A$, the
sequence $q_n(s_n)$ converges to $q(s)$, then the sequence of sets
$\bigcap_{s \in A_n} H(s,q_n(s))$ converges to $\bigcap_{s \in A}
H(s,q(s))$ in the Pompeiu-Hausdorff distance---provided either the
limit set is the closure of its interior, or it is a singleton and
the sets in the sequence are nonempty.
\end{Thm}

We illustrate the use of this theorem on two examples. In the first,
we take $q_n(s) = q(s) = Q(p,s)$; the theorem then says that the
successive approximations, $D_{A_1}(p) \supseteq D_{A_2}(p) \supseteq
\dots$, approach $D_{A}(p)$ in the Pompeiu-Hausdorff distance.
Typically, $A_n$ are finite, while $A = \mathbb{S}^{d-1}$; the only
requirements is that the directional quantiles $Q(p,s)$ depend on $s$
in a continuous way---for instance, $P$ satisfies the assumptions of
Theorem~\ref{t:quacont}. The second example leads to a proof of
consistency, similar to those given by \citet{HeWan97}, of
$\hat{D}_n(p)$ to $D(p)$, when $\hat{D}_n(p)$ arise via applying the
definition of directional quantile envelopes to empirical
distributions that converge weakly almost surely to the sampled
population distribution $P$ (under suitable scheme, say, of
independent sampling); the required assumptions are those of
Theorem~\ref{t:quacont}, continuous or bounded support of $P$, and the
nondegeneracy of the limit $D(p)$ (in general we cannot guarantee that
$\hat{D}_n(p)$ are nonempty). The Skorokhod representation then yields
random variables $X_n$ converging almost surely to random variables
$X$, such that the laws of $X_n$ and $X$ are the corresponding
empirical distributions and $P$, respectively; Theorem~\ref{t:quacont}
then implies the convergence assumption required by
Theorem~\ref{t:hausdorff}.

To obtain some idea about the magnitude of the approximation error, we
may proceed as follows (for simplicity, we limit our scope to the
two-dimensional setting). Let $d \in \partial D$. The directions of
all tangents of a convex set $D$ at $d$ generate a convex cone,
$T_D(d)$. Let $c_D(d)$ be the maximal cosine between its two
directions, the cosine of the maximal angle between two extremal
normalized directions in $T_D(d)$,
\begin{equation*}
c(d) = \sup\, \biggl\{ \frac{s\trans t}{\|s\| \|t\|} \colon s,t \in
T_D(d) \biggr\} = \sup\, \bigl\{ s\trans t \colon s,t \in T_D(d)
\cap \mathbb{S}^{d-1} \bigr\}.
\end{equation*}
In fact, this cosine is the same as the maximal cosine of the
directions in the normal cone $N_D(d)$; see \citet{RocWet98},
Chapter 6. We can see that $c_D(d) \leq 1$, the equality holding if
and only if $T_D(d)$ consists of single direction---when $D$ has a
unique tangent at $d$. Let $ \kappa_D = \sup_{d \in \partial D}
\sqrt{{2}/(1+c(d))}, $ the reciprocal of the cosine of the
\emph{half} of the maximal angle between directions in the tangent
cone. Apparently, $\kappa_D \geq 1$, the equality holding true for
smooth $D$. On the other hand, $\kappa_D$ can be equal to $+\infty$
for the degenerate $D$, the sets with empty interior.
\begin{Thm}
\label{t:consist} Let $A \subseteq \mathbb{S}^1$ be a set of
directions, and let $\hat{q}(s)$ and $q(s)$ be two functions on $A$.
Suppose that both $\hat{D}=\bigcap_{s \in A} H(s,\hat{q}(s))$ and $D
= \bigcap_{s \in A}H(s,q(s))$ are nondegenerate; then both
$\kappa_{\hat{D}}$ and $\kappa_D$ are finite and
\begin{equation*}
d\bigl( \hat{D}, D \bigr) \leq \max
\{\kappa_{\hat{D}},\kappa_{\vphantom{\hat{D}}D}\} \sup_{s \in A}
|\hat{q}(s)-q(s)|,
\end{equation*}
where $d$ denotes the Pompeiu-Hausdorff distance.
\end{Thm}

\medskip
\setcounter{equation}{0} 
\noindent {\bf 7. Directional quantile envelopes beyond simple
  location setting}

So far, our methodology was demonstrated in the simple location
setting, the situations
when there are no covariates and the estimation is performed via the
application of the quantile operators to empirical distributions. In
this section, we show how the directional definition can be used in
more sophisticated constructions.

\begin{figure}[!t]
\begin{center}
\includegraphics[width=.45\textwidth, height=.25\textheight]{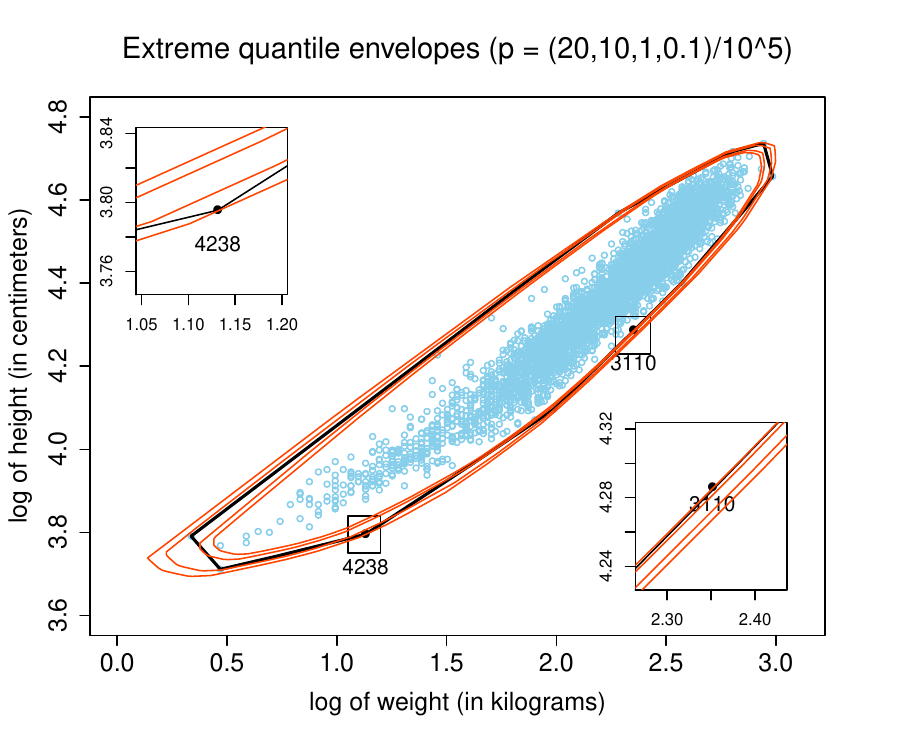}
\includegraphics[width=.45\textwidth, height=.25\textheight]{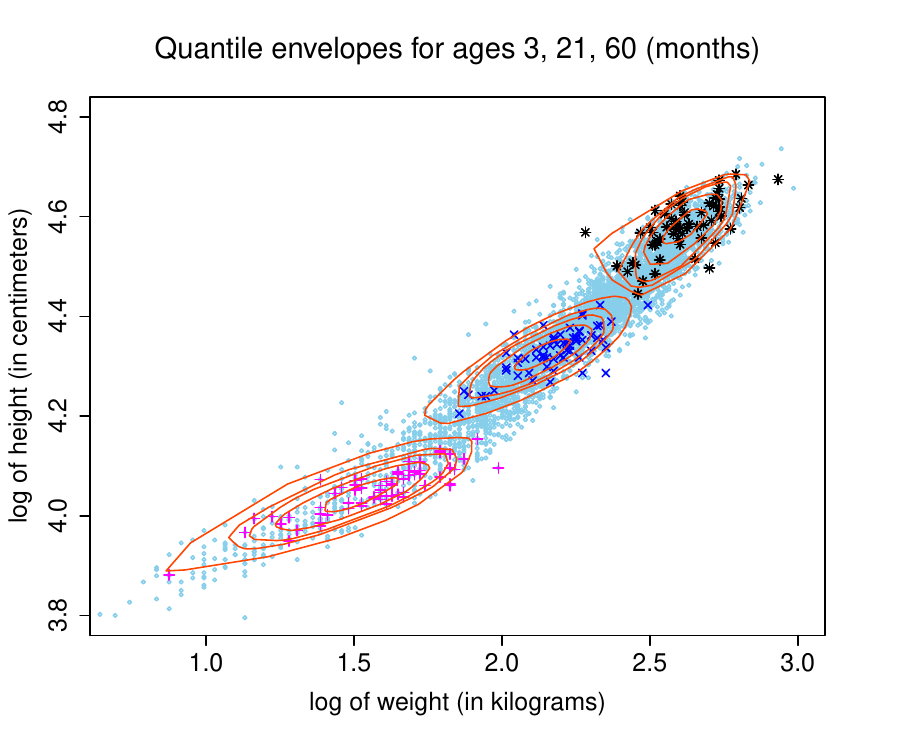}
\caption{\it Left panel: directional extremal quantiles, derived from
  the corresponding univariate analogs, and the convex hull, the
  empirical extremal quantile. Right panel: imagine an animation in
  which the directional quantile envelopes slowly ascend upward along
  the data cloud, demonstrating the dependence on the increasing
  covariate, age.}
\label{f:12}
\end{center}
\end{figure}

The first application is to extreme quantiles. It is apparent that
this type of analysis calls for other than empirical estimators of
population quantiles: if , say, $100$ observations are available, then
their maximum, the $p$-th empirical quantile for any $p > 0.99$, may
not be found satisfactory for estimating a threshold with exceedance
probability less than, say, $0.001$. This is well known, and this
paper neither proposes any new take on the subject, nor sides with any
of the approaches that can be found in \citet{BeiGoeTeu04},
\citet{ReiTho07}, \citet{Res07}, and the references given there. Our
only message here is that once the problem is satisfactorily handled
in the univariate case, the directional philosophy allows for an
immediate extension to the multivariate setting. The result can be
seen in the left panel of Figure~\ref{f:12}. The estimated extreme
quantiles, for $p=10^{-6}$, $10^{-5}$, $10^{-4}$, and $2 \times
10^{-4}$ are confronted with the convex hull of the data, the
empirical estimate for any $p \leq (2.33)10^{-4}$. The plot seems to
provide some information about the extent of extremality of the points
labeled by 3110 and 4238; a closer inspection reveals that 3110 lies
on the ($2 \times 10^{-3}$)-th directional quantile envelope, while
4238 on the ($10^{-6}$)-th one. The real worth of this information
crucially depends on the properties of the estimates of extreme
quantiles in the univariate case; in this particular case we chose
those we found in the R package \verb|evir| \citep{McNSte07}, due to
their nonparametric flavor and ready availability of the
implementation.



\begin{figure}[!b]
\begin{center}
\includegraphics[width=.245\textwidth,height=.19\textheight]{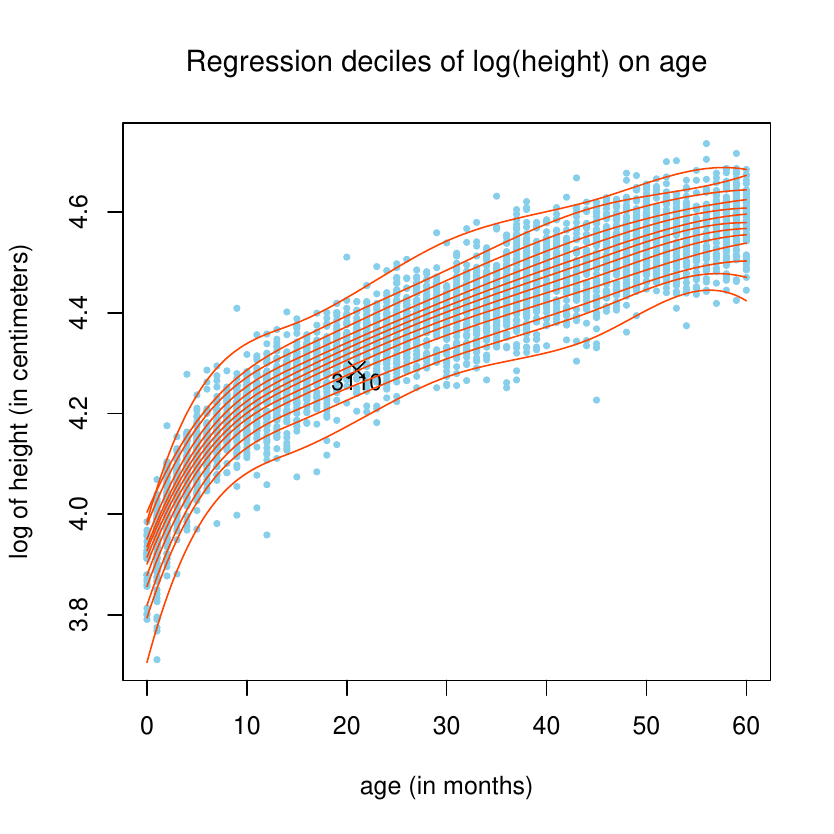}
\includegraphics[width=.245\textwidth,height=.19\textheight]{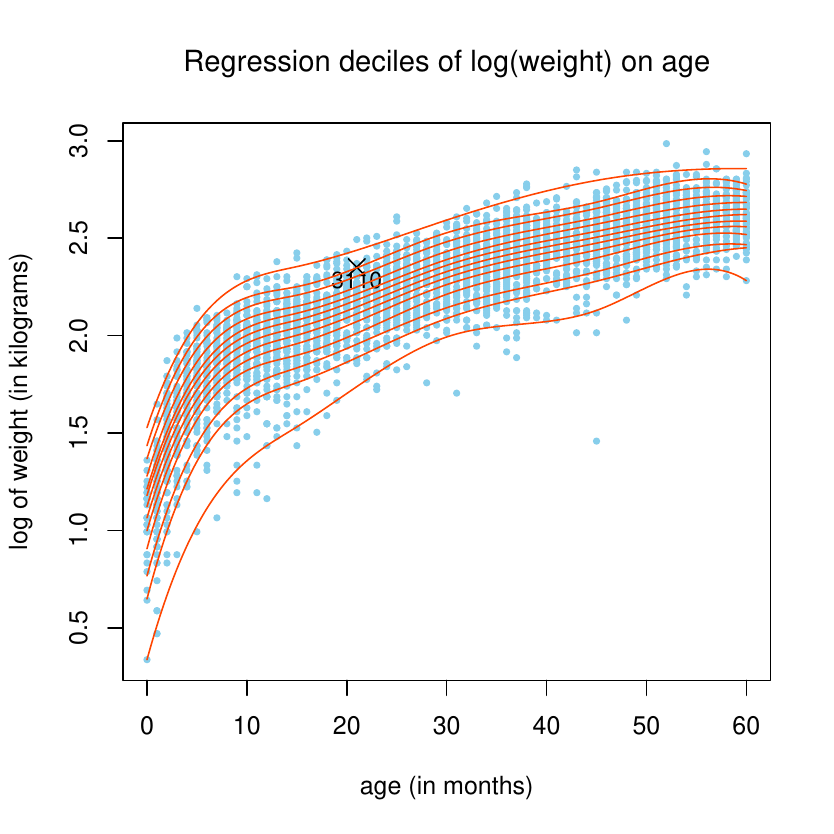}
\includegraphics[width=.245\textwidth,height=.19\textheight]{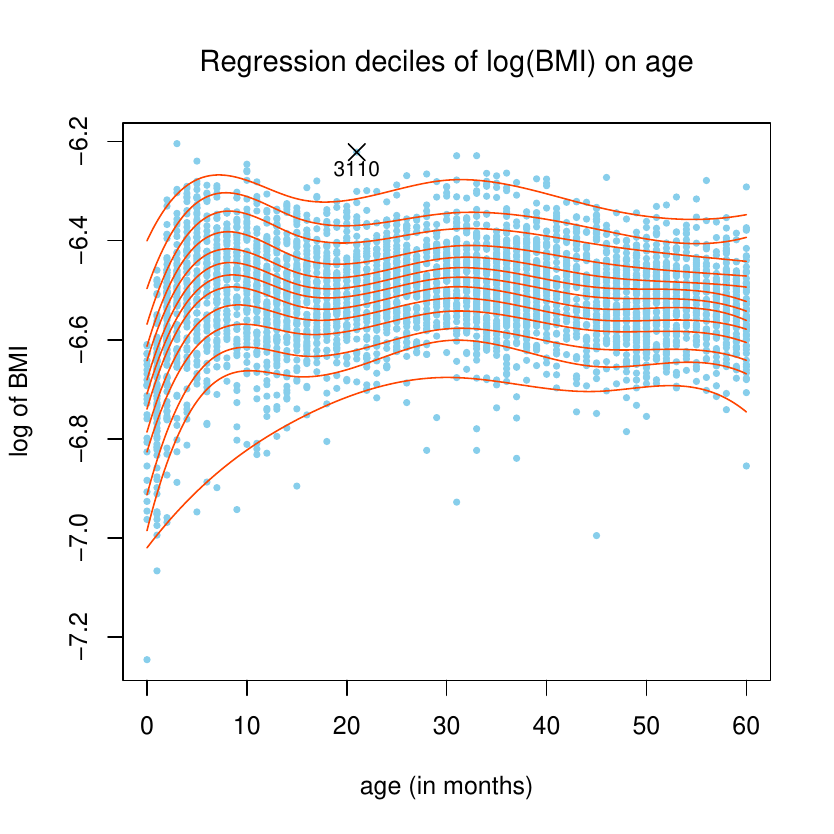}
\includegraphics[width=.245\textwidth,height=.19\textheight]{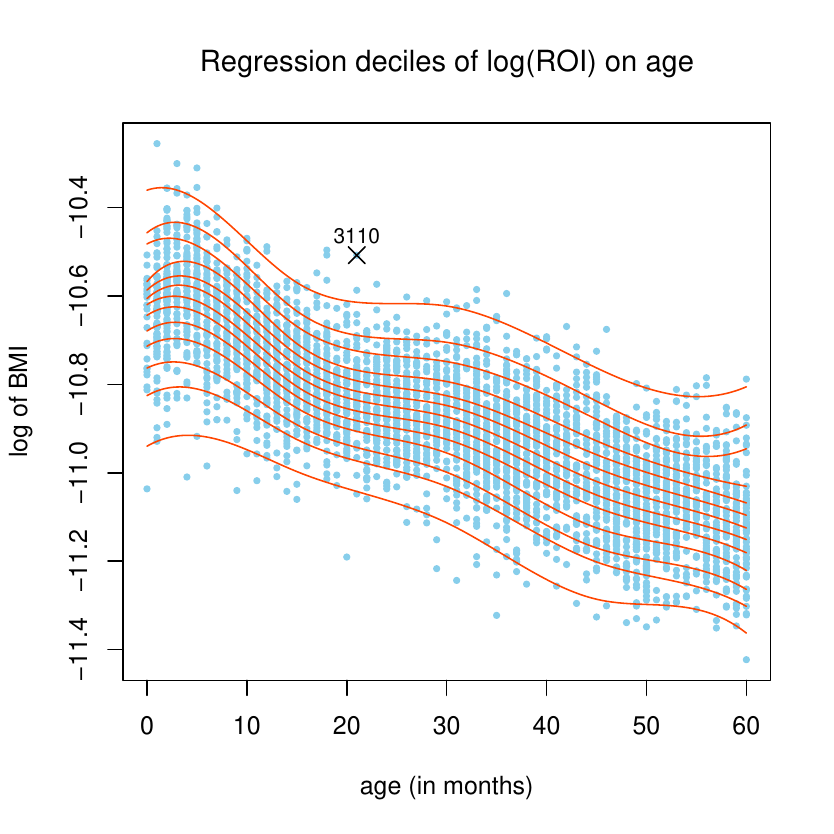}
\caption{\it The ``growth charts'', quantiles of various linear
  combinations of the primary variables, regressed on the covariate,
  age.}
\label{f:13}
\end{center}
\end{figure}

Our second application are bivariate growth charts. While the relevant
proposal of \citet{Wei08}, roughly characterized as quantile
regression in polar coordinates, is conceptually capable of delivering
interpretable contours, its practical application is plagued by its
considerable dependence on the underlying nonparametric regression
methodology, implemented in the R package \verb|cobs| by
\citet{NgMae06}, and by the necessity to select the origin of the
coordinate system---somewhat ad hoc task, which nevertheless seriously
influences the result. These and other shortcomings (in particular,
the tendency of the estimated contours to intersect themselves) led to
the pursuit of the approach we propose here; some precursor ideas were
outlined already by \citet{SalZam06}. It should be noted that our way
of indexing is different from that of \citet{Wei08}, so her approach
may be, after all, viewed as complementary rather than alternative to
ours.

Figure~\ref{f:13} shows the deciles of several projections of the
vector response, consisting of the logarithm of weight and height,
regressed on the covariate, the age in months. While such ``growth
charts'' facilitate a lot of useful insights, the user may like to
confront them with a directional perspective---in a related
covariate-dependent context. Such a desire stumbles upon the
inevitable fact that our graphical universe is two-dimensional;
animations and interactive graphics are certainly possible, but in the
traditional setting we can merely opt for plotting of the directional
quantiles for some fixed value(s) of the covariate---as in the right
panel of Figure~\ref{f:12}, which shows the predicted envelopes for
three values of the age (selected so that the resulting envelopes do
not overplot, rather than pursuing any other objective). The
highlighted datapoints represent the subjects with the particular age.
If we computed directional quantile envelopes from these points
separately, the resulting contours would be rougher, and would vary
from one value of age to another; the contours given in the right
panel of Figure~\ref{f:12} borrow strength from other ages,
constructing quantile envelopes from a number of quantile regressions,
like those seen in Figure~\ref{f:13}.

Once again, our focus here is on how quantile regression blends into
directional quantile philosophy; from this perspective, our rendering
of nonparametric quantile regression rather avoided than explored
potential challenges, and we refer to \citet{Koe05} and references
there for the fine aspects of the methodology. In view of
Theorem~\ref{t:affine}, our main concern was whether the estimates are
translation (regression) and scale equivariant, to yield affine
equivariant envelopes---which is true if the fits in
Figures~\ref{f:13} and the right panel of \ref{f:12} are obtained via
regression splines, the methodology used by \citet{WeiPerKoe05} in
their paper on growth charts. We used the automated knot selection
furnished by the R package \verb|splines| \citep{Tea07}, and fitted
quantile regressions by the R package \verb|quantreg| \citep{Koe07}.
The smoothing parameter was selected by eyeballing the plots included
in Figure~\ref{f:13}, and then adopting a universal smoothing
parameter for all directions in the right panel of Figure~\ref{f:12}.
We are aware of the possible shortcomings in certain engineering
details---for instance that unlike in our situation, one can easily
imagine data exhibiting more signal-to-noise in certain directions
than in others, the fact that would have to be reflected in varying
smoothing parameters; we hope to address these problems in future
research, as well as explore alternative possibilities for
nonparametric quantile regression in the construction of growth
charts.

\medskip
\setcounter{equation}{0} 
\noindent {\bf 8. Final remarks}

This paper is a shortened version of the preprint of
\citet{KongMizera08}---whose exposition was, unfortunately, perhaps
too subtle or somewhat misconceived, so that it brought a lot of
misunderstanding from its readers and referees; we are indebted to all
of them for alerting us to this defect. The present version drops any
mention of impasses like ``quantile biplots'' (so that they will not
be confused with the concepts that we really champion); space
considerations led us to omit the discussion of fine aspects of growth
charts; finally, we do not discuss any computational details, as those
were rendered obsolete by the meanwhile development---we hope to
address computational aspects in a separate publication.

We would like to stress that our objective was not to propose any
``multivariate quantile'' generalization of the univariate concept,
akin to those reviewed in \citet{Ser02}---we believe that rather than
such generalizations, a way of condensing and presenting the
information about the quantiles of univariate functions of the data is
needed, so that the specific and well-recognizable \emph{meaning} of
quantiles, as expounded in this paper, transcribes into multivariate
context. It is possible that existing pursuits of ``multivariate
quantiles'' aimed, at least nominally, at similar objectives; thence
it may be understandable when our efforts are viewed in a certain
vicinity of theirs.

We conclude that directional quantile envelopes---which are,
essentially, depth contours---are a possible way to condense
directional quantile information, the information carried by the
quantiles of projections. In typical circumstances, they allow for
relatively faithful and straightforward retrieval of the directional
quantile information; the methodology offers straightforward
probabilistic interpretations, and the estimated quantile envelopes
are affine equivariant under mild equivariance assumptions on the
estimators of directional quantiles. Most importantly, the directional
interpretation can be adapted to elaborate frameworks requiring more
sophisticated quantile estimation methods than evaluating quantiles
for empirical distributions, including estimation of extreme quantiles
and directional quantile regression.


\medskip
\noindent {\bf Acknowledgment}

We are indebted to Ying Wei for turning our attention to multivariate
growth charts, as well as for many insights in \citet{Wei08}; and to
Roger Koenker for valuable discussions. The directional approach to
depth contours was pioneered in the unpublished master thesis of
Beno\^it Laine, as reported by \citet{Koe05}---in, however, quite
significantly more complicated version fitting not directional
quantiles, but directional quantile regressions. Another important
forerunners were \citet{SalZam06}. This research was supported by the
Natural Sciences and Engineering Research Council of Canada; some of
the results originate from the doctoral dissertation of
\citet{Kong09}.

\par

\smallskip
\setcounter{equation}{0} 
\noindent {\bf Appendix. Proofs}

We define the $p$-th \emph{directional quantile set} to be the
quantile set of the corresponding projection:
\begin{equation*}
\mathcal{Q}(p,s) = \mathcal{Q}(p,s,X) = \mathcal{Q}(p, s\trans X).
\end{equation*}

\noindent{\bf  Proof of Theorem~\ref{t:quacont}} Since
quantile sets are bounded intervals, it is sufficient to prove the
convergence of their endpoints to $\inf \mathcal{Q}(p,s\trans X) =
\inf \{u \colon \mathbb{P}[s\trans X \leq u ] \geq p\}$ and $\sup
\mathcal{Q}(p,s\trans X) = \sup \{u \colon \mathbb{P}[s\trans X \geq
  u] \leq (1-p)\}$.

Suppose that the support of $X$ is bounded. Let $q = \inf
\mathcal{Q}(p,s\trans X)$; we have that $\mathbb{P}[s\trans X \leq q
] \geq p$ and $\mathbb{P}[s\trans X \leq
  q-\epsilon ] < p$. If the support of the distribution of $X$ is
bounded, we have $\| X \| \leq M$ almost surely; by the Schwarz
inequality, $|(s -s_n)^T X | \leq M \| s -s_n \|$ and therefore
\begin{equation*}
\label{!one} p \leq \mathbb{P}[s\trans X \leq q ] =
\mathbb{P}[s_n\trans X \leq q - (s-s_n)\trans X] \leq
\mathbb{P}[s_n\trans X \leq q + M \| s -s_n \|],
\end{equation*}
which means that $\inf \mathcal{Q}(p,s_n^TX) \leq q + M \| s -s_n
\|$. In a similar fashion, we obtain that $\inf
\mathcal{Q}(p,s_n^TX)\geq q - M \| s -s_n \|-\epsilon$, due to
$\mathbb{P}[s_n\trans X \leq q - M \| s -s_n
\|-\epsilon] \leq \mathbb{P}[s\trans  X \leq q  - \epsilon] < p.
$
Letting $\epsilon
\rightarrow 0$, we obtain
$
q - M \| s -s_n \| \leq \inf \mathcal{Q}(p,s_n\trans X) \leq q + M
\| s -s_n \|,
$
and therefore $\inf \mathcal{Q}(p,s_n\trans X) \rightarrow \inf
\mathcal{Q}(p,s\trans X)$, and thus also $Q(p,s_n,X_n)$ to
$Q(p,s,X)$. The convergence of $\sup \mathcal{Q}(p,s_n\trans X)
\rightarrow \sup \mathcal{Q}(p,s\trans X)$ is proved analogously.

If the support of the distribution of $X$ is contiguous, then all
directional quantile sets in the limit are singletons.
Pompeiu-Hausdorff convergence then follows from the ``outer
convergence'' of quantile sets in the sense of \citet{RocWet98}, see
also \citet{MizVol02}: any limit point, $x$, of any sequence $x_n
\in \mathcal{Q}(p,s_n,X_n)$ lies in $\mathcal{Q}(p,s,X)$. This can
be easily seen in an elementary way, observing that $x_n \in
\mathcal{Q}(p,s_n,X_n)$ entails
$$
p \leq \limsup_{n\to\infty} \mathbb{P}[s_n\trans X_n \leq x_n] \leq
\mathbb{P}[s\trans X \leq x]$$
and
$$
1-p \leq \limsup_{n\to\infty} \mathbb{P}[s_n\trans X_n \geq x_n]
\leq \mathbb{P}[s\trans X \geq x].
$$
Since under the contiguous support assumption the quantiles are
unique, this second part of the theorem holds true for every
quantile version.

\noindent{\large\bf Proof of Theorem~\ref{t:basic}} If $y \in D(p)$,
then $y \in H(p,s)$ for every $s \in \mathbb{S}^{d-1}$ and thus $P(\{x
\colon s\trans x\geq s\trans y\})\geq p$ for all $s \in
\mathbb{S}^{d-1}$; therefore $d(x) \geq p$. Conversely, if $d(y)\geq
p$, then for every $s \in \mathbb{S}^{d-1}$ we have $P(\{x \colon
s\trans x\geq s\trans y\})\geq p$. It follows that $s\trans y \geq
Q(p,s)$ and thus $y \in H(p,s)$. Hence $y \in D(p)$. As already
mentioned, this theorem is true only for the ``inf'' definition, other
quantile versions give smaller envelopes.

\noindent{\bf  Proof of Theorem~\ref{t:linglong}} See
\cite{KongZuo10}.

\noindent{\bf  Proof of Theorem~\ref{t:elliptic}} By
rotational invariance, the directional quantile envelopes of any
circularly symmetric distribution are circles; since elliptic
distributions are those that can be transformed to the circular
symmetric ones by an affine transformation, the theorem follows from
their affine equivariance (and holds true for any quantile version).

\noindent{\bf  Proof of Theorem~\ref{t:smooth}} See
\cite{KongZuo10}.

\noindent{\bf  Proof of Theorem~\ref{t:affine}} Let $B$ be a
nonsingular matrix and $b$ a vector. First, we verify that the
transformation rule for the supporting halfspace of the directional
quantile: for every $s \in \mathbb{S}^{d-1}$ and every $p \in
(0,1)$,
\begin{equation}
\label{!transrule} H(B^{\ast}s/\| B^{\ast}s \|, Q(p,s,BX + b) ) = B
H(s,Q(p,s,X)) + b,
\end{equation}
where $B^{\ast} = (B^{-1})\trans$. If  $B$ is orthogonal,
then $B^{\ast} = B$, and if $B$ is diagonal (more generally,
symmetric), then $B^{\ast} = B^{-1}$. Indeed, the equation satisfied
by $x$ in $BH(s,(Q,p,s,X))$,
$
s\trans \left( B^{-1}x \right) \leq Q(p,s,X),$
~is equivalent to~
$((B^{-1})\trans s)\trans x = (B^{\ast} s)\trans x \leq Q(p,s,X).
$
The norm of $s$ is one, but not necessarily that of $B^{\ast}s$;
therefore, we divide both sides by $\| B^{\ast} s \|$,
\begin{equation}
\label{!bbbq} (1/\|B^{\ast} s \|) (B^{\ast} s)\trans x \leq
(1/\|B^{\ast} s\|) Q(p,s,X).
\end{equation}
By the scale equivariance of the quantile operator, and by the
relationship $Q(p, s, AX) = Q(p, A\trans s, X)$, which follows
directly from the definition, we obtain that the right-hand side of
\eqref{!bbbq} equals
$
Q(p,s,X/\|B^{\ast} s\|) = Q(p,s/\|B^{\ast} s\|, X) = Q(p,B^{\ast}
s/\|B^{\ast} s\|, BX).
$
 Since the transformation $BX+b$ is one-to-one, the
transformed intersection of halfspaces is the intersection of
transformed halfspaces. Therefore, the transformed directional
quantile envelope is by \eqref{!transrule}
\begin{equation*}
\bigcap_{s\in \mathbb{S}^{d-1}} \left( B H(p,s,X) + b \right) =
\bigcap_{s\in \mathbb{S}^{d-1}} H(p, B^{\ast}s/\| B^{\ast}s \|, BX +
b).
\end{equation*}

The proof is concluded by observing that $s \mapsto B^{\ast}s /\|
B^{\ast}s \|$, where $ B^{\ast} = (B^{-1})\trans$, is a one-to-one
transformation of $\mathbb{S}^{d-1}$ onto itself---as can be seen by
the direct verification involving its inverse, $t \mapsto B\trans t
/ \| B\trans t \|$. The proof is the same for any quantile version.

\noindent {\bf  Proof of Theorem~\ref{t:hausdorff}} To prove
convergence with respect to Pompeiu-Hausdorff distance, we exploit
the following facts. First, the sequence $\bigcap_{s \in A_n}
H(s,q_n(s))$, together with the limit $\bigcap_{s \in A} H(s,q(s))$
is contained in a bounded set, starting from some $n$. This follows
from the fact that sets $A_n$ are approaching a dense set in $A$,
and the latter is not contained in any halfspace whose boundary
contains the origin; therefore this property is shared by $A_n$
starting from some $n$, which means that
$
\bigcap_{s \in A_n} H(s,\inf_{k \geq n} q_n(s))
$
is the desired bounded set. For uniformly bounded sequences, the
convergence in Pompeiu-Hausdorff distance follows from the
convergence in Painlev\'e-Kuratowski sense; see \citet{RocWet98},
4.13. The latter means that a general sequence of sets $K_n$
converges to $K$ if (i) every limit point of any sequence $x_n \in
K_n$ lies in $K$; (ii) every point from $K$ is a limit of a sequence
$x_n \in K_n$. See also \citet{MizVol02}.

For sequences of closed sets with ``solid'' limits, sets that are
closures of their interior, the Painlev\'e-Kuratowski convergence
follows from the ``rough'' convergence, defined by
\citet{LucSalWet94} to require (i) together with (ii)' every limit
point of every sequence $y_n \in (\operatorname{int} K_n)^c$ is in
$(\operatorname{int} K)^c$. See also \citet{LucTorWet93}. That is,
one can replace outer and inner convergence requirement of
Painlev\'e-Kuratowski definition by two outer convergences, the
original one, and the other one for ``closed complements''.

Suppose that $y \in \operatorname{int} K$. Then $y$ belongs to all
but finitely many $K_n$; otherwise, there would be a subsequence
$n_i$ such that $y \in (\operatorname{int} K_{n_i})^c$, and by the
modified version of (ii)', $y \in (\operatorname{int} K)^c$. Hence,
every $y$ from the relative interior of $K$ is a limit of an
(eventually constant) sequence $y_n \in K_n$. To obtain (ii) for
every $x \in K$, consider a sequence $y_k$ of points from (nonempty)
$\operatorname{rint} K$ such that $y_n \to y$; the desired sequence
$x_n$ is then obtained by a ``diagonal selection'': for every $y_k$,
there is $n_k$ such that $y_k \in K_i$ for every $i \geq k$; set
$x_n = y_k$ for every $n_k \leq n < n_{k+1}$.

Thus, it is sufficient to prove (i) and (ii)'. Suppose that $x$ is a
limit point of a sequence $x_n \in \bigcap_{s \in A_n} H(s,q_n(s))$.
Then there is a subsequence such that $s_n\trans x_n \geq q_n(s_n)$
for every $s_n \in A_n$; every $s \in A$ is a limit of a sequence
$s_n \in A_n$, therefore the assumptions of the theorem imply that
$s\trans x \geq q(s)$; hence $x \in \bigcap_{s \in A} H(s,q(s))$.
This proves (i). This proves theorem for the singleton case, since
then the Painlev\'e-Kuratowski convergence is implied by (i) once
the sets in the sequence are nonempty.

Suppose now that $x$ is a limit point of a sequence $x_n \in \left(
\operatorname{int} \bigcap_{s \in A_n} H(s,q_n(s)) \right)^c$, that
is, a limit of some subsequence of $x_n$. Every such $x_n$ satisfies
$s_n\trans x_n \leq q_n(s_n)$ for some
$s_n \in A_n$. By the compactness of $A$, there is $s \in A$ that is
a limit of a subsequence of $s_n$; passing to the limit along the
appropriate subsequences, we obtain that $s\trans x \leq q(s)$, by
the assumptions of the theorem. This means that $x \in \left(
\operatorname{int} \bigcap_{s \in A} H(s,q(s)) \right)^c$.

\noindent{\bf  Proof of Theorem~\ref{t:consist}} As $\hat{D}$
and $D$ are compact convex sets, we have $d(\hat{D},D) = d(\partial
\hat{D},\partial D)$. Let $\epsilon = \sup_{s \in A} |\hat{q}(s) -
q(s)|$; we will show that for any $x \in \partial D$, $d(x,
\partial\hat{D} ) \leq \kappa_{D}\epsilon$. Let
$\tilde{q}(s)=q(s)-\epsilon$ and $\tilde{D} = \bigcap_{s \in A}
H(s,\tilde{q}(s))$.

For simplicity, we assume that $\tilde{D}$ is also nondegenerate. We
have that $\tilde{D}\subseteq\hat{D}$, and also $\tilde{D} \subseteq
D$, the latter set being congruent to $\tilde{D}$. If $\kappa_{D}(x)
> 1$, then $x$ is a vertex of $D$. Since $d(x,\tilde{x}) =
\kappa_{D}(x)\epsilon$, where $\tilde{x}$ is the corresponding
congruent vertex in $\partial \tilde{D}$, it follows that
$d(x,\partial \tilde{D}) \leq \kappa_{D}\epsilon$. When
$\kappa_{D}(x)=1$, then, by Theorem 24.1 of \citet{Roc96}, there
exists a sequence $x_n \ne x$, $x_n \in\partial D$, such that $x_n
\rightarrow x$ and $s_n \rightarrow s$, $\kappa_{D}(x_n)=1$, where
$s_n$ and $s$ are the directions of the tangent lines passing
through $x_n$ and $x$, respectively. There are two possibilities.

If there is $N$ such that $s_n=s$ for any $n>N$, then there must be
two points, denoted by $y_1$ and $y_2$, in $\partial
H(s,q(s))\cap\partial D$ such that $\kappa_{D}(y_1)>1$ and
$\kappa_{D}(y_2)>1$. That is, $y_1$ and $y_2$ are two vertices of
$D$ and there is no other vertex between $y_1$ and $y_2$ of $D$.
Suppose that $\tilde{y}_1$ and $\tilde{y}_2$ are points congruent to
them on $\tilde{D}$; then $\tilde{y}_1$ and $\tilde{y}_2$ are two
vertices of $\tilde{D}$ and there is no other vertex between
$\tilde{y}_1$ and $\tilde{y}_2$ of $\tilde{D}$ as well. In other
words, we have a trapezoid with vertices $y_1$, $y_2$, $\tilde{y}_1$
and $\tilde{y}_2$ and $x$ lies on one of the bases. A simple
geometric calculation then shows the existence of a point, $y$,
lying on the base constructed by $\tilde{y}_1$ and $\tilde{y}_2$,
such that $d(x,y)\leq \max
\{\kappa_{D}(y_1),\kappa_{D}(y_2)\}\epsilon$, that is, $d(x,\partial
\tilde{D})\leq \kappa_{D}\epsilon$.

Suppose that there is an infinite subsequence of $s_n$ such that
$s_n\neq s$. Let $\partial \tilde{D}$ by $\tilde{x}_n$ and
$\tilde{x}$ be the congruent counterparts of $x_n$ and $x$,
respectively; let $s_n$ and $s$ be the corresponding directions. Let
$y_n = \partial H(s_n,q(s_n)) \cap \partial H(s,q(s))$ and
$\tilde{y}_n = \partial \tilde{H}(s_n,q(s_n)) \cap \partial
\tilde{H}(s,q(s))$. We have that $y_n\rightarrow x$,
$\tilde{y}_n\rightarrow \tilde{x}$, and
$d(y_n,\tilde{y}_n)=\sqrt{2}\epsilon/\sqrt{1+s_n^Ts}$. As
$d(y_n,\tilde{y}_n)\rightarrow d(x,\tilde{x})$ and
$\sqrt{2}\epsilon/\sqrt{1+s_n^Ts}\rightarrow \epsilon$, we arrive to
$d(x,\tilde{x})=\epsilon$, which means $d(x,\partial \tilde{D})\leq
\kappa_{D}\epsilon$ again.

Taking into account that $\tilde{D}\subseteq \hat{D}$, we obtain
that $d(x,\partial \hat{D})\leq \kappa_{D}\epsilon$, for any $x\in
\partial D$. The theorem follows form this and the symmetric
inequality, $d(x,\partial D)\leq \kappa_{\hat{D}}\epsilon$ holding
true for any $x\in \partial \hat{D}$, which can established in an
analogous way.


\bibliographystyle{apalike}
\bibliography{KongMizera}


\vskip .65cm \noindent Linglong Kong and Ivan Mizera\\
Department of Mathematical and Statistical Sciences\\
University of Alberta\\
CAB 632, Edmonton, Alberta, T6G 2G1 Canada \vskip 2pt \noindent
E-mails: lkong@ualberta.ca and mizera@stat.ualberta.ca\vskip 2pt


\end{document}